\documentclass[final,5p,times,twocolumn]{elsarticle}
\usepackage{mathrsfs}

\usepackage{amssymb}
\usepackage{amsmath}
\usepackage{esvect}
\begin{document}

\begin{frontmatter}


%%       addressline={},
%%       city={},
%%       postcode={},
%%       state={},
%%       country={}}

%\title{Large Recoil Symmetric Compton Scattering Theory for Nuclear Physics application}
%\title{Symmetric Compton Scattering: A Novel Method to Generate Gamma Rays from Low-Energy Electrons}
\title{Full Inverse Compton Scattering: Total Transfer of Energy and Momentum from Electrons to Photons}%for Nuclear Physics Applications
%using Deep Recoil inverse Compton Scattering

%%       addressline={},
%%       city={},
%%       postcode={},
%%       state={},
%%       country={}}
%%       addressline={},
%%       city={},
%%       postcode={},
%%       state={},
%%       country={}}

\author[INFN-MI]{L. Serafini}
\author[INFN-MI,UNIMI]{V. Petrillo}
\author[INFN-MI]{S. Samsam \corref{cor1}}
\cortext[cor1]{Email address: sanae.samsam@mi.infn.it}

\affiliation[INFN-MI]{organization={INFN-Milano and LASA},%Department and Organization
      addressline={Via G. Celoria 16}, 
      city={Milan},
      postcode={20133}, 
      state={},
      country={Italy}}

\affiliation[UNIMI]{organization={University of Milan},%Department and Organization
      addressline={Via G. Celoria 16}, 
      city={Milan},
      postcode={20133}, 
      state={},
      country={Italy}}

%\affiliation[Elettra-Sincrotrone]{organization={Elettra-Sincrotrone},%Department and Organization
      %addressline={Basovizza}, 
      %city={Trieste},
      %postcode={34149}, 
      %state={},
      %country={Italy}}

\begin{abstract}

 In this article we discuss a peculiar regime of Compton Scattering that assures the maximum transfer of energy and momentum from free electrons propagating in vacuum to the scattered photons. We name this regime Full Inverse Compton Scattering (FICS) because it is characterized by the maximum and full energy loss of the electrons in collision with photons: up to 100 $\%$ of the electron kinetic energy is indeed transferred to the photon. In the case of relativistic electrons, characterized by a large Lorentz factor ($\gamma \gg 1$), FICS regime corresponds to an incident photon energy equal to $\frac{m_e c^2}{2}$, i.e. approximately 255.5 keV. We interpret such an astonishing result as FICS being the time reversal of direct Compton Scattering of very energetic photons (of energy much greater than $m_e c^2$) onto atomic electrons. Although the cross section of Compton scattering is decreasing with the energy of the incident photon, making the process less probable with respect to other reactions (pair production, nuclear reactions, etc) when high energetic photons are bombarding a target, the kinematics straightforwardly implies that the back-scattered photons would have an energy reaching asymptotically $\frac{m_e c^2}{2}$. FICS is instead the unique suitable working point in Compton scattering for achieving the total transfer of (kinetic) energy exactly from the electron to the photon. Experiencing transitions from the initial momentum to zero in the laboratory system, in FICS the electron is also subject to very large negative acceleration; this fact can lead to possible experiments of sensing the Unruh temperature and related photon bath. On the other side of the energy dynamic range, low relativistic electrons can be completely stopped by moderate energy photons (tens of keV), leading to full exchange of temperature between electron clouds and photon baths. Cosmic gamma ray sources can be affected in their evolution by this peculiar FICS regime of Compton scattering.
 
\end{abstract}



\begin{keyword}
Inverse Compton Scattering \sep Electron energy loss \sep Cosmic Gamma-ray sources \sep Energy transfer between electrons and photons \sep Unruh radiation.



\end{keyword}

\end{frontmatter}


\section{Introduction}
\label{sec:intro}

The study of Compton Scattering dates back to 1923, when Arthur H. Compton first observed the scattering of X-rays by atomic electrons in solid targets, leading to the discovery of what is now known as the Compton effect \cite{PhysRev.21.483}. This groundbreaking work provided direct evidence for the particle nature of light, supporting the emerging quantum theory and earning Compton the Nobel Prize in Physics in 1927. Since then, Compton Scattering has become a fundamental process in quantum electrodynamics (QED), extensively used to probe the structure of matter and the behavior of high-energy particles \cite{Lawson2014ComptonSF}.

In the realm of high-energy particle physics, Compton Scattering has long been a cornerstone for understanding the interaction between electrons and photons. Traditionally, this scattering process results in a partial transfer of energy and momentum from an electron to a photon, a phenomenon well-described by QED. However, in this study, we explore a remarkable and unconventional regime of Compton Scattering, termed Full Inverse Compton Scattering (FICS). In this regime, an electron propagating in vacuum can transfer 100$\%$ of its kinetic energy to an incident photon of proper energy. This represents an extension of our previous work, where we analyzed various transition points and regimes of inverse Compton scattering (ICS), including the significant transition point where the electron is brought to rest by colliding with a photon of energy ($m_e c^2/2$) (deep recoil regime) \cite{serafiniFICS}. By focusing on this specific scenario, we aim to delve deeper into the energy loss and transfer efficiency in FICS compared to conventional Compton Scattering.
Despite the decreasing cross-section of Compton Scattering with increasing photon energy making the process less probable compared to other high-energy interactions such as pair production and nuclear reactions the kinematics dictate that fully back-scattered photons from highly energetic photon collisions with atomic electrons would asymptotically reach energies of 255 keV. This surprising result highlights the unique characteristics of FICS, distinguishing it inside the dynamical range of Compton Scattering, from direct kinematics towards inverse kinematics, as the continental divide of the two regimes. As a matter of fact in direct Compton Scattering the electron is mainly back or highly scattered, while in Inverse Compton Scattering the electron keeps propagating close to the prolongation of its incident trajectory path even after scattering. In FICS the electron just stops after scattering with the incident photon in the ideal case of a fully back-scattered photon, but in any case the electron after scattering is taken down to kinetic energies smaller than $m_ec^2/2$ for all photons that are back-scattered within an angle $\theta=1/\gamma$, which represents a substantial fraction of scattering events in the condition of FICS with relativistic electrons.

FICS stands out as the only interaction regime in vacuum where the kinetic energy of an electron can be wholly transferred to a photon, resulting in maximum negative acceleration of the electron. This acceleration can reach ultimate extreme levels, scaling like the energy of the incident electron.  Its effects open potential avenues for experimental investigations into sensing the Unruh temperature, sometimes called the Davies–Unruh temperature, a phenomenon predicted by quantum field theory in accelerated frames as the effective temperature experienced by a uniformly accelerating detector in a vacuum field \cite{RevModPhys.80.787}. Moreover, at the lower end of the energy spectrum, even non-relativistic electrons can be fully halted by photons with moderate energies (tens of keV), enabling a complete thermal exchange between electron clouds and photon baths. This peculiar scattering regime may significantly influence the evolution of cosmic gamma-ray sources, adding a new dimension to our understanding of astrophysical phenomena and the energy dynamics in extreme environments.
In this comprehensive study, we delve deeply into the FICS transition, rigorously comparing the energy loss and transfer efficiency in fully Inverse Compton Scattering versus conventional Compton Scattering. By analyzing the kinematic conditions and the energy exchange dynamics, we aim to uncover the fundamental principles governing this regime. Additionally, the study explores the implications of FICS in various high-energy scenarios, providing insights into how such interactions could manifest in both terrestrial and cosmic settings.
%to be verified
One of the most promising applications of FICS lies in the field of plasma physics, particularly in enhancing the efficiency of electron trapping and heating. In our previous work, we demonstrated the feasibility of trapping electrons in magnetic bottles for plasma heating applications \cite{Serafini:2024yah}. By utilizing the unique properties of FICS, electrons can be stopped immediately after their interaction with photons, resulting in an immediate and complete transfer of kinetic energy. This advancement not only enhances the control over the electron population within the plasma but also improves the overall efficiency of the heating process, potentially leading to more effective methods for achieving and maintaining high-temperature plasma.
Furthermore, this regime could be used to create highly sensitive and precise gamma-ray detectors. By ensuring that electrons can be completely stopped by incoming photons, detectors can achieve higher energy resolution and efficiency. This application would be particularly useful in astrophysics for detecting and analyzing cosmic gamma-ray sources.

%Furthermore, the ability to fully stop electrons with moderate energy photons suggests new possibilities for managing electron behavior in various technological applications, from particle accelerators to radiation shielding. The insights gained from studying FICS could pave the way for novel approaches in controlling and utilizing electron-photon interactions, with broad implications for both fundamental research and practical applications.

\section{Full Inverse Compton Scattering Theory}
\label{sec:FICStheory}

Understanding Compton scattering regimes is crucial in various scientific disciplines, with applications that depend on the specific initial photon and electron energies, as well as the nature of the experiment. In this manuscript, we focus on the FICS transition point where a maximum transfer of energy and momentum from the electron to the photon takes place.

Since the electron cannot materially vanish, the maximum energy transferable to the photon is all its kinetic energy $T_e = (\gamma-1)m_e c^2$. This point is highly significant for numerous applications, such as gamma-ray production for sustainability purposes. For instance, to generate a gamma-ray beam while minimizing power and energy expenditure on the electron beam, it is essential to maximize the energy extracted from the electron beam and transferred to the gamma-ray beam. Otherwise, the remaining energy would be dissipated as thermal energy.

The key to this transition is setting an energy and momentum budget, ensuring that the total energy and momentum of the system remain invariant after scattering.
This process can be analyzed starting from the conservation of energy and momentum conservation principles: for sake of simplicity we assume that the electron-photon collision is head-on, so that their transverse momentum components are vanishing, as the total transverse momentum of the system (we assume z-axis as the initial propagation axis of the electron, travelling towards positive z's, and the photon, propagating towards negative z's).  %, while the the one of the scattered photon is negative.
\begin{equation}
\label{eq:ConsevationE}
    E_e + E_{ph} = E_{e}^{\prime} + E_{ph}^{\prime}
\end{equation}

\begin{equation}
\label{eq:ConservationP}
    p_{ze} + p_{zph} = p_{ze}^{\prime} + p_{zph}^{\prime}
\end{equation}

\begin{equation}
\label{eq:ConservationP3.eq3}
    0 = p_{xe}^{\prime} + p_{xph}^{\prime}
\end{equation}

where $E_{e}=\gamma m_e c^2$ and $p_{ze}= \gamma m_e v_e $ are the energy and momentum of the electron before scattering. $E_{ph}$ and $p_{ph}$ are the energy and momentum of the photon before scattering.
$\gamma$ is the Lorenz factor, $m_e$ is the electron rest mass, $v_e =\beta c$ is the electron velocity and $\beta$ its dimensionless velocity. We also assume that (x,z) is the plane of scattering.% Attenzione ho aggiunto la Eq.3, rinumerare tutte le equazioni.

Building on the principles outlined in references \cite{PhysRevAccelBeams.21.030701,serafini2023symmetric,serafiniFICS, PhysRevAccelBeams.20.080701}, and by applying the conservation laws of total energy and momentum, the energy of a photon scattered at an angle $\theta$ relative to the z-axis can be determined using the following formula:

\begin{equation}
\label{eq:Eph_compton}
    E_{ph}^{\prime} (\theta)= \frac{(1+\beta)\:E_{ph} E_e}{(1-\beta \: cos \: \theta) E_e+ (1+cos \: \theta)\: E_{ph}}
\end{equation}

\begin{figure}[!ht]
    \centering
    \includegraphics[width=\columnwidth]{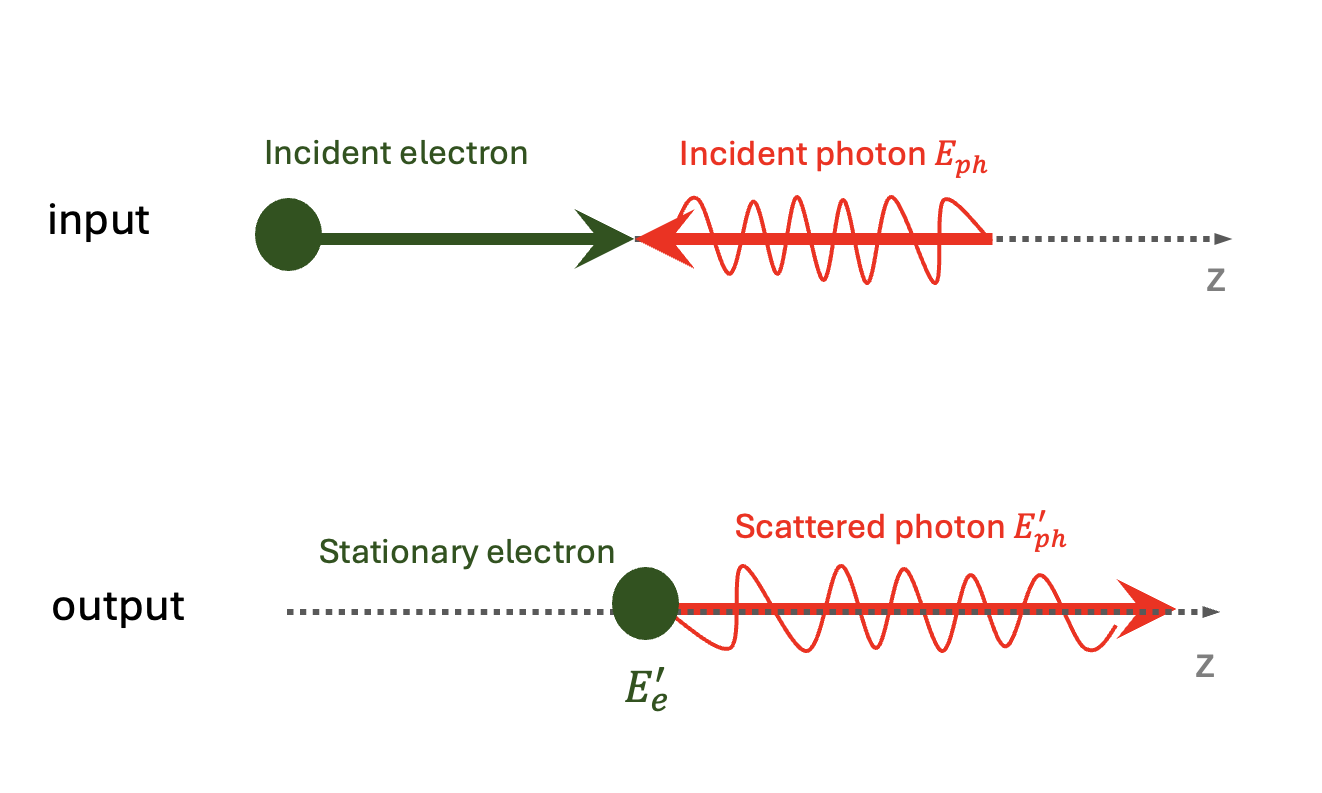}
    \caption{Top: Head-on collision between an electron (green) and a photon (red). Bottom: Fully inverse Compton scattering with a stationary electron, where the incident photon is fully scattered at an angle $\theta=0$ and gains higher energy. In this process, the stationary electron transfers all its kinetic energy to the scattered photon.}
    \label{fig:scheme_ficsheadon}
\end{figure}

Using Eq.\ref{eq:ConsevationE} and \ref{eq:ConservationP3.eq3}, and assuming that maximum energy/momentum transfer between the electron and the photon occurs when the scattering angle in Eq.\ref{eq:ConservationP3.eq3} is $\theta =0$, we can easily derive the expression of the kinetic energy of the electron after scattering, $T_{e}^{\prime}$, as follows:

\begin{equation}
\label{eq:Te'_fics}
    T_{e}^{\prime} = (\gamma-1)m_e c^2 + E_{ph} - \frac{(1+\beta) \: \gamma \: E_{ph} m_e c^2}{(1-\beta)\:\gamma m_e c^2 + 2 E_{ph}}
\end{equation}
%D
with $\gamma=1/\sqrt{1-\beta^2}$ is the electron Lorentz factor. This equation holds true for any arbitrary values of the initial energies and momenta of the colliding electron and photon. It is quite straightforward to see that the kinetic energy of the electron after scattering, $T_{e}^{\prime}$, is asymptotically going down to 0 when $\gamma$ tends to infinity, independently on the value of the incident photon energy $E_{ph}$.
This property was firstly recognized in Refs. \cite{Follin_thesis} and \cite{PhysRev.73.449}, thoroughly discussed in Ref. \cite{serafiniFICS} (see Eq.7). When the recoil factor $X$ , defined as $X=\frac{4E_eE_{ph}}{(m_0c^2)^2}$ , approaches infinity, the electron asymptotically tends to transfer its all kinetic energy to the scattered photon.
Here instead we focus on FICS working point, searching for an exact solution of Eq.\ref{eq:Eph_compton} so that $T_{e}^{\prime}$=0. We find that it exists a particular value of $E_{ph}$ for which $T_{e}^{\prime}$=0 for any value of the incident electron energy $E_e$ (or its corresponding Lorentz factor $\gamma$). The solution of $T_{e}^{\prime} =0$ is given by:

\begin{equation}
\label{eq:E_fics1}
    E_{ph} = E_{ph}^{FICS} = \frac{m_e c^2}{2} (1-\gamma+\beta \gamma)
\end{equation}

If the energy of the incident photon $E_{ph}$ is given by Eq.6, and the photon is back-scattered at $\theta=0$ then an electron of energy $E_e=\gamma m_e c^2$ will stop after scattering, releasing all its kinetic energy to the photon. If its Lorentz factor gamma is much larger than 1, i.e. in case of relativistic electrons, the energy of the incident photon $E_{ph}^{FICS}$ requested to have FICS is just $E_{ph}^{FICS}$ = $m_e c^2/2$, i.e. almost 255.5 keV. 
An alternative simpler way to understand the unique properties of FICS can be understood through the principle of energy and momentum conservation imposing explicitly the FICS condition. In this process, the longitudinal component of the electron's momentum after scattering becomes zero, and its energy is reduced to its rest mass energy when the scattering angle $\theta$ is zero. This scenario is illustrated in Fig. \ref{fig:scheme_ficsheadon}.
By rewriting Eqs \ref{eq:ConsevationE} and \ref{eq:ConservationP} in terms of energy, we can determine the energies of both the photon and electron before and after scattering, which is valid for any value of $\gamma$ and $\beta$, imposing explicitly:

\begin{equation}
\label{eq:ConsevationE_fics}
    \gamma m_e c^2 + E_{ph} = m_e c^2 + E_{ph}^{\prime}
\end{equation}
\begin{equation}
\label{eq:ConservationP_fics}
    \beta \gamma m_e c^2 - E_{ph} = 0 + E_{ph}^{\prime}
\end{equation}

Solving these 2 equations for $E_{ph}^{\prime}$ and $E_{ph}$, we derive the scattered photon and electron energies for FICS, which prove the steps above to obtain Eq.\ref{eq:E_fics1}:

\begin{equation}
\label{eq:consEph}
    E_{ph}^{FICS} = \frac{m_e c^2}{2} (1 -(1-\beta)\:\gamma)
\end{equation}

\begin{equation}
\label{eq:consEphp}
    E_{ph}^{\prime} = \frac{m_e c^2}{2} ((1+\beta)\: \gamma -1)
\end{equation}

\begin{equation}
\label{eq:eep}
    E_{e}^{\prime} = m_e c^2
\end{equation}

where $E_{ph}^{FICS}$ is the incident photon energy requested for FICS (note that Eq.9 is equal to Eq.6). In this case, and by using Eq. \ref{eq:eep}, we can easily demonstrate that $T_{e}^{\prime} = E_{e}^{\prime}-m_e c^2=0$, which means that the electrons are totally at rest.
In the relativistic case, where $\gamma \gg 1$ and therefore, $\beta$ is approximately $1- \frac{1}{{2\gamma^2}}$, Eqs. \ref{eq:consEph} and \ref{eq:consEphp} can be rewritten as follow:

\begin{equation}
\label{eq:Ephrelativistic}
    E_{ph} = \frac{m_e c^2}{2} (1- \frac{1}{{2\gamma}})
\end{equation}

\begin{equation}
\label{eq:Ephprelativistic}
    E_{ph}^{\prime} = m_e c^2 (\gamma - \frac{1}{2}- \frac{1}{4\gamma})
\end{equation}

clearly showing that any ultra-relativistic electron is stopped by a 255.5 keV ($m_e c^2/2$) photon in case they undergo Compton scattering with head-on collision and fully back-scattered photon.

\begin{figure}[!ht]
    \centering
    \includegraphics[width=\columnwidth]{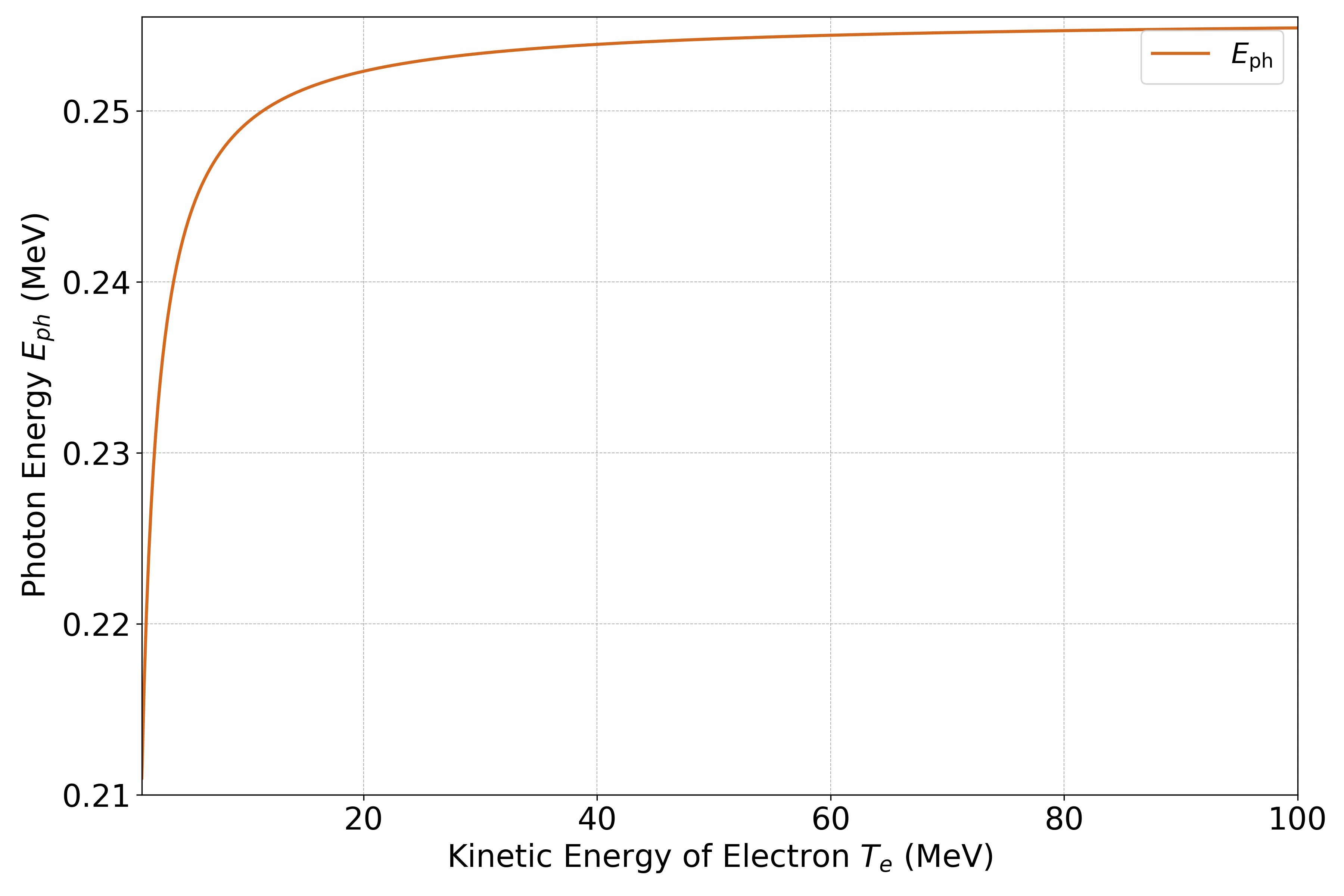}
    \caption{FICS in relativistic case. Incident photon $E_{ph}$ as a function of the incident electron kinetic energy $T_e$ in the MeV range}
    \label{fig:FICSrcase}
\end{figure}

\begin{figure}[!ht]
    \centering
    \includegraphics[width=\columnwidth]{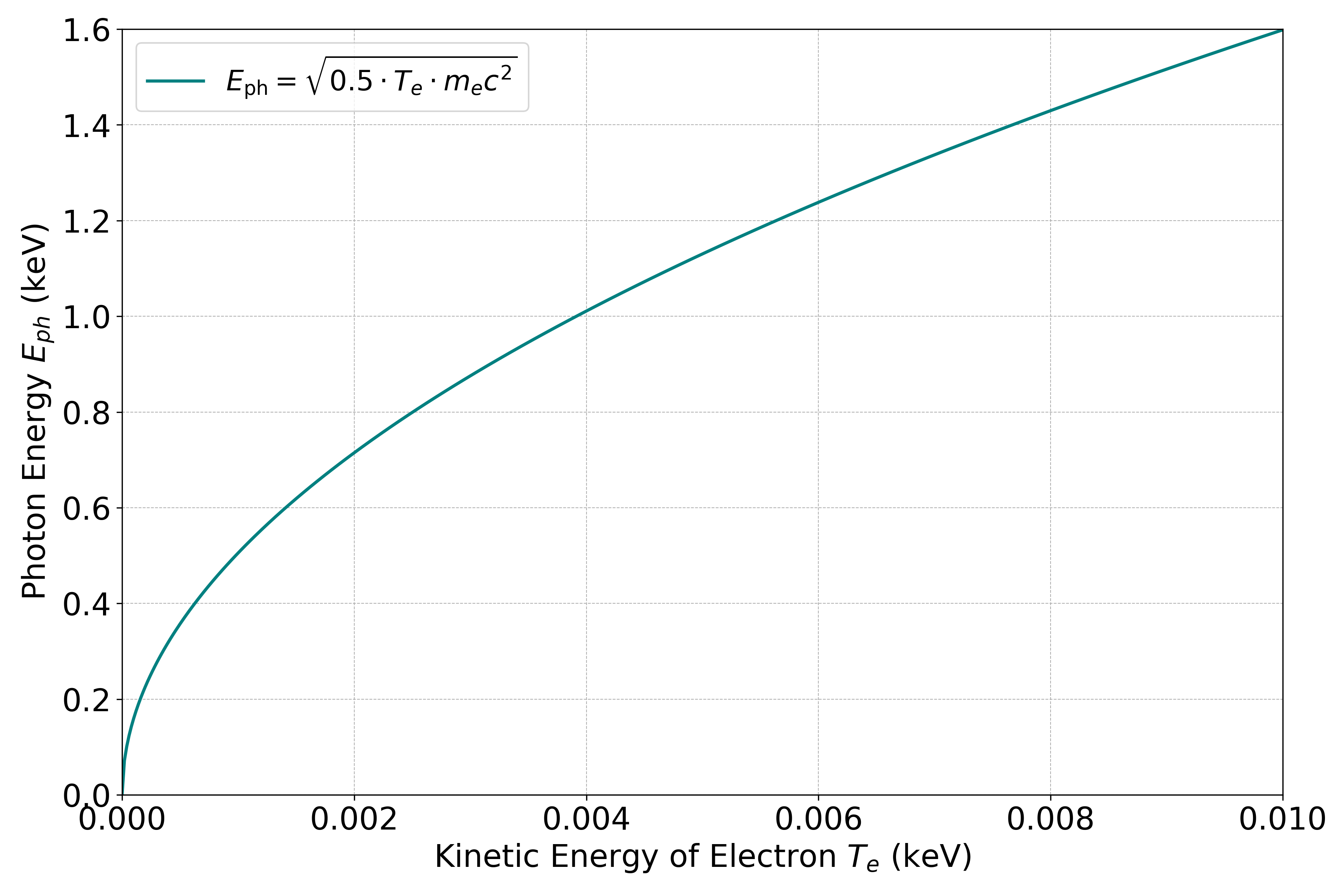}
    \caption{Incident photon energy $E_{ph}$ as a function of the incident electron's kinetic energy $T_e$ in the non-relativistic region, presented in keV. This plot covers $T_e$ up to 500 keV, highlighting the photon energy behavior at lower kinetic energies.}
    \label{fig:ficsnonRv}
\end{figure}

%add more details about the plot
Fig. \ref{fig:FICSrcase} depicts the photon energy as function of the electron kinetic energy demonstrating how the value of $E_{ph}$ for FICS rapidly approaches 255.5 keV, providing a clear visualization of the photon energy behavior at higher kinetic energy. In the other hand, Fig. \ref{fig:ficsnonRv}, focuses on the non-relativistic region, presenting $E_{ph}$ for lower kinetic energy values. 

In this last case, where $\beta \ll 1$ and $\gamma \approx 1+\frac{1}{2} \beta^2$, Eqs. \ref{eq:consEph} and \ref{eq:consEphp} can expressed in the forms:

\begin{equation}
\label{eq:consEph_lr}
    E_{ph} = \beta \:\frac{m_e c^2}{2}
\end{equation}

\begin{equation}
\label{eq:consEphp_lr}
    E_{ph}^{\prime} \approx \beta\: \frac{m_e c^2}{2}
\end{equation}

\begin{equation}
\label{eq:ee_lr}
    E_e = m_e c^2 \:(1+\frac{1}{2} \beta^2)
\end{equation}

where the energy of the scattered electrons $E_{e}^{\prime}$ remains equal to its rest mass since the electron stops after collision.

In the non-relativistic regime, characterized by $\beta \ll 1$, the comparison between FICS and SCS (Symmetric Compton Scattering) where the energy/momentum transfer between photons and electrons is balanced reveals significant insights. For SCS with non-relativistic electrons, the energy of the incident photon $E_{ph}=\beta m_e c^2$, as detailed in Eq. 22 in Ref. \cite{serafini2023symmetric}. Conversely, to perform FICS in the non-relativistic regime, the required incident photon energy is $E_{ph}=\beta \frac{m_e c^2}{2}$. This implies that FICS necessitates a photon with half the energy required for SCS.

\begin{figure}[!ht]
    \centering
    \includegraphics[width=\columnwidth]{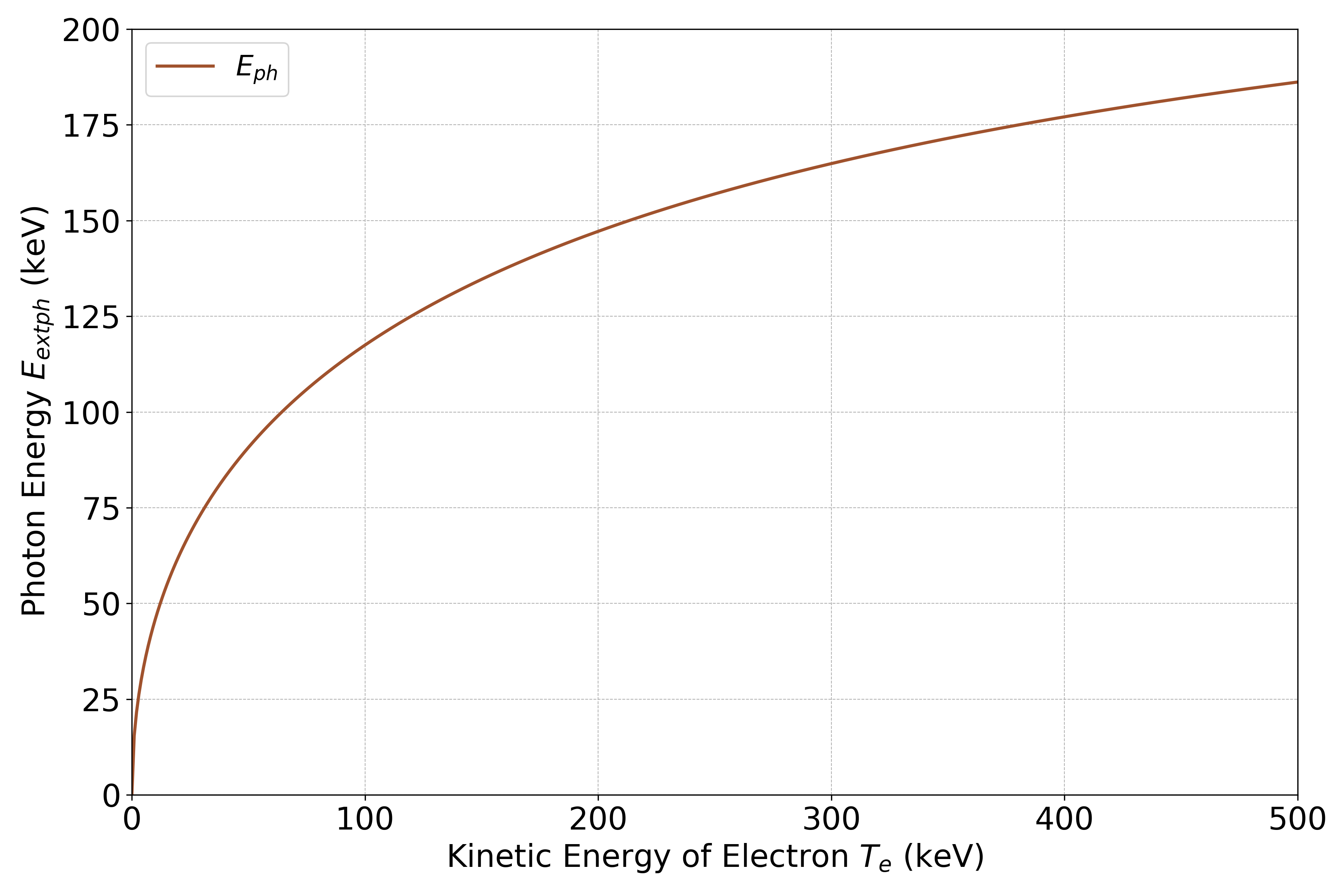}
    \caption{Incident photon energy $E_{ph}$ as a function of the incident electron kinetic energy $T_e$  in the non-relativistic regime. The plot covers  $T_e$ from 0 to 10 eV and $E_{ph}$ up to 1.6 keV.}
    \label{fig:ficsnonrvTe10ev}
\end{figure}

 Always in the non-relativistic case, regime, where the scaling law for the incident photon energy required to achieve FICS is $E_{ph}= \sqrt{T_e \frac{m_e c^e}{2}}$ which is only valid for $T_e \ll 10 \:$keV. For instance, to stop an electron with a kinetic energy of 1 eV, a photon energy of approximately 505 eV is necessary. Fig. \ref{fig:ficsnonrvTe10ev} illustrates this relationship for $T_e$ up to 10 eV.

\section{Energy Loss and Transfer Efficiency in Compton Scattering: Direct Compton vs. Fully Inverse Compton Scattering}

In analyzing the energy dynamics of particle collisions, particularly between high-energy photons and electrons, the concepts of energy loss and transfer efficiency become crucial. This examination delves into the comparative analysis of both DC (Direct Compton) and FICS regimes, focusing on their respective energy transfer coefficients and loss factors.

In Direct Compton scattering, the energy transfer coefficient is defined as the ratio of the incident photon energy $E_{ph}$ to the kinetic energy $T_e= (\gamma-1)m_e c^2$ of the scattered electron. The electron's total energy $E_e$ comprises its kinetic energy and rest mass energy ($E_e$= $T_e + m_e c^2$), but only the kinetic energy is available for transfer during the collision, as the electron cannot disappear.

The energy loss factor for DC is given by:
\begin{equation}
    \label{eq:lossfacDC}
    E_{\mathrm{loss,DC}} = \frac{T_{e}^{\prime}}{E_{ph}} = 1 - \frac{1}{2 X_{DC}}
\end{equation}

where $X_{DC}$= $\frac{4E_{ph}}{m_e c^2}$ is the electron recoil factor of DC interactions \cite{serafini2023symmetric}. This energy loss factor approaches 1 asymptotically as the recoil $X_{DC}$ tends to infinity, indicating that higher incident photon energies result in more efficient energy transfer to the electron.
In contrast, the FICS point presents a different dynamic. Here, the energy loss factor is expressed as:
\begin{equation}
    \label{eq:lossfacFICS}
    E_{\mathrm{loss,FICS}} = \frac{E_{ph}^{\prime}}{T_{e}} = \frac{m_e c^2(\gamma-1/2)}{m_e c^2(\gamma-1)}= \frac{\gamma-1/2}{\gamma-1}
\end{equation}

Remarkably, this ratio is greater than 1, implying that the electron can transfer more energy to the photon than its initial kinetic energy. This occurs because the scattered photon benefits from both the electron's kinetic energy and the incident photon's energy (approximately $\frac{m_e c^2}{2}$=255.5 keV). If we instead calculate the ratio between the scattered photon energy and the incident electron energy, we find:
\begin{equation}
    \label{eq:ratioEph/Ee}
    \frac{E_{ph}^{\prime}}{E_e} \approx \frac{m_e c^2(\gamma-1/2)}{\gamma m_e c^2}= 1-{\frac{1}{2\gamma}}
\end{equation}

This ratio is always slightly less than 1, indicating that the photon energy after scattering is almost equal to the total electron energy.
The analysis reveals that FICS is more effective than DC in transferring energy and momentum between the photon and electron. The higher efficiency of FICS makes it a significant consideration in high-energy particle interactions and warrants inclusion in the ongoing research on Fully Inverse Compton Scattering.

The FICS point where the electron transfers its entire kinetic energy to the photon, can be examined as well using Eq. 5 from our previous paper \cite{serafini2023symmetric}, where we find the stationary point (maximum) of the energy transfer function as a function of the recoil $X$.
In order To find this stationary point, we take the derivative of the energy transfer function with respect to $X$ and set it to zero. The condition for the maximum is given by $X=2\gamma$. At this maximum, we can define the ratio $\frac{E_{ph}^{\prime}}{E_{tot}}$ as follow:

\begin{equation}
    \label{eq:rationE/Etot}
    \frac{E_{ph}^{\prime}}{E_{tot}} = \frac{2\gamma}{(1+1/2\gamma)(1+ 2\gamma)} 
\end{equation}

This ratio rapidly approaches 1 as $\gamma$ becomes much larger than 1 ($\gamma \gg 1$), Thus, at high $\gamma$ values, FICS is the point where the energy of the scattered photon is almost equal to the total energy of the electron. This indicates that the scattered photon can effectively gain nearly all the electron's energy, making FICS highly efficient in energy transfer. 
Eq. \ref{eq:rationE/Etot} can also be reformulated in terms of $X$ and  the asymmetry factor A (A=$\gamma^2 - \frac{X}{4}$)to provide a clearer perspective on the energy transfer efficiency:

\begin{equation}
    \frac{E_{ph}^{\prime}}{E_{tot}} = \frac{X(4A+X)}{(1+X)(4A+2X)} 
\end{equation}

This transformation presents a neat representation of the relationship between the incident photon energy, the electron's total energy, and the energy of the scattered photon.
While the previous derivation using the electron's kinetic energy $T_e$ s more precise, it is important to note that when $\gamma \gg 1$ , the electron's kinetic energy and its total energy are nearly equal. Therefore, for high-energy electrons, the distinctions between using kinetic energy and total energy in calculating energy transfer efficiency become negligible.
%Scatterd electrons after FICS transition:
\section{Kinetic Energy and Momentum Analysis of Scattered Electrons in FICS with Small Angle Perturbations}

\begin{figure}[!ht]
    \centering
    \includegraphics[width=\columnwidth]{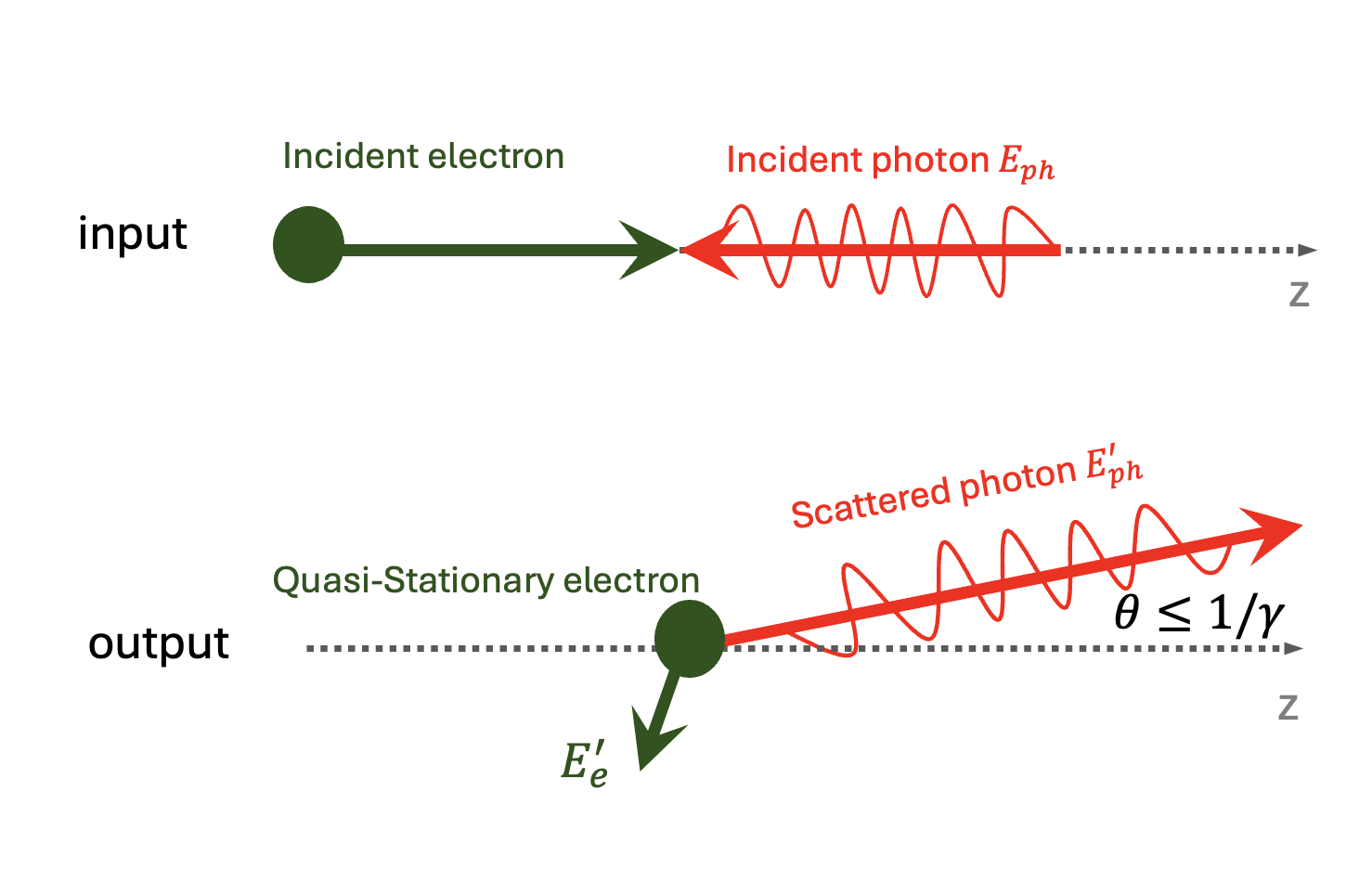}
    \caption{Top: Head-on collision between an electron (green) and a photon (red). Bottom: Inverse Compton scattering with a quasi-stationary electron, where the incident photon is scattered at a small angle $\theta$ and gains energy. In this process, the electron transfers almost all its kinetic energy to the scattered photon.}
    \label{fig:scheme_theta}
\end{figure}

In this section we generalize the study of FICS to  generic photon scattering angles $\theta$, in order to evaluate both the residual non-zero kinetic energy of the electron after scattering ($T_{e}^{\prime}$) and the fraction of electrons that remain quasi-stationary after scattering, with respect to the total scattering events over the full solid angle. The electron kinetic energy within a scattering angle $\theta$ can be written as, generalizing Eq.\ref{eq:Te'_fics} to generic $\theta$ values:

\begin{equation}
\label{eq:Te'_fics_theta}
    T_{e}^{\prime} = T_e + E_{ph}^{FICS} - \frac{(1+\beta)\:\gamma\: E_{ph}^{FICS} m_e c^2}{(1-\beta\: cos\: \theta)\: \gamma m_e c^2 + (1+cos\: \theta) \: E_{ph}^{FICS}}
\end{equation}

Assuming relativistic electrons $\gamma \gg 1$ and considering only scattering angles smaller than $\theta \approx \frac{1}{\gamma}$ , the kinetic energy $T_e^{\prime}$ of the electron simplifies in:

\begin{equation}
\label{eq:Te'_fics_theta_gamma}
    T_{e}^{\prime} = \frac{m_e c^2}{2} \gamma^2 \theta^2
\end{equation}

For small angles $\theta$, where $cos \theta \approx 1- \frac{\theta^2}{2}$ but considering generic values of gamma we find that $T_e^{\prime}(\theta=1/\gamma)$ simplifies to the following form:

\begin{equation}
\label{eq:Te'_fics_1/g}
    T_{e}^{\prime}(\theta=1/\gamma) = m_e c^2 \frac{4\gamma^2 - 4\gamma -1}{8\gamma^2 - 4\gamma }
\end{equation}
As $\gamma$ approaches infinity,  $T_e^{\prime}$ converges to:
\begin{equation}
\label{eq:Te'_fics_1/g2}
    T_{e}^{\prime}(\theta=1/\gamma) \xrightarrow{\gamma \rightarrow \infty} \frac{m_e c^2}{2}
\end{equation}

In agreement with Eq. \ref{eq:Te'_fics_theta_gamma}, this result shows that all electrons undergoing FICS with photons back-scattered within the scattering angle $\theta=1/\gamma$ will have a residual kinetic energy after scattering smaller than $m_ec^2/2$, which corresponds to a $\beta$ smaller than 0.75. In other words, all these electrons will be non-relativistic after scattering, therefore they will be subject to a large velocity drop from before to after scattering, of at least 0.25$\times c$ or larger (up to c for electrons undergoing FICS with photons fully back-scattered at $\theta$=0).

\begin{figure}[!ht]
    \centering
    \includegraphics[width=\columnwidth]{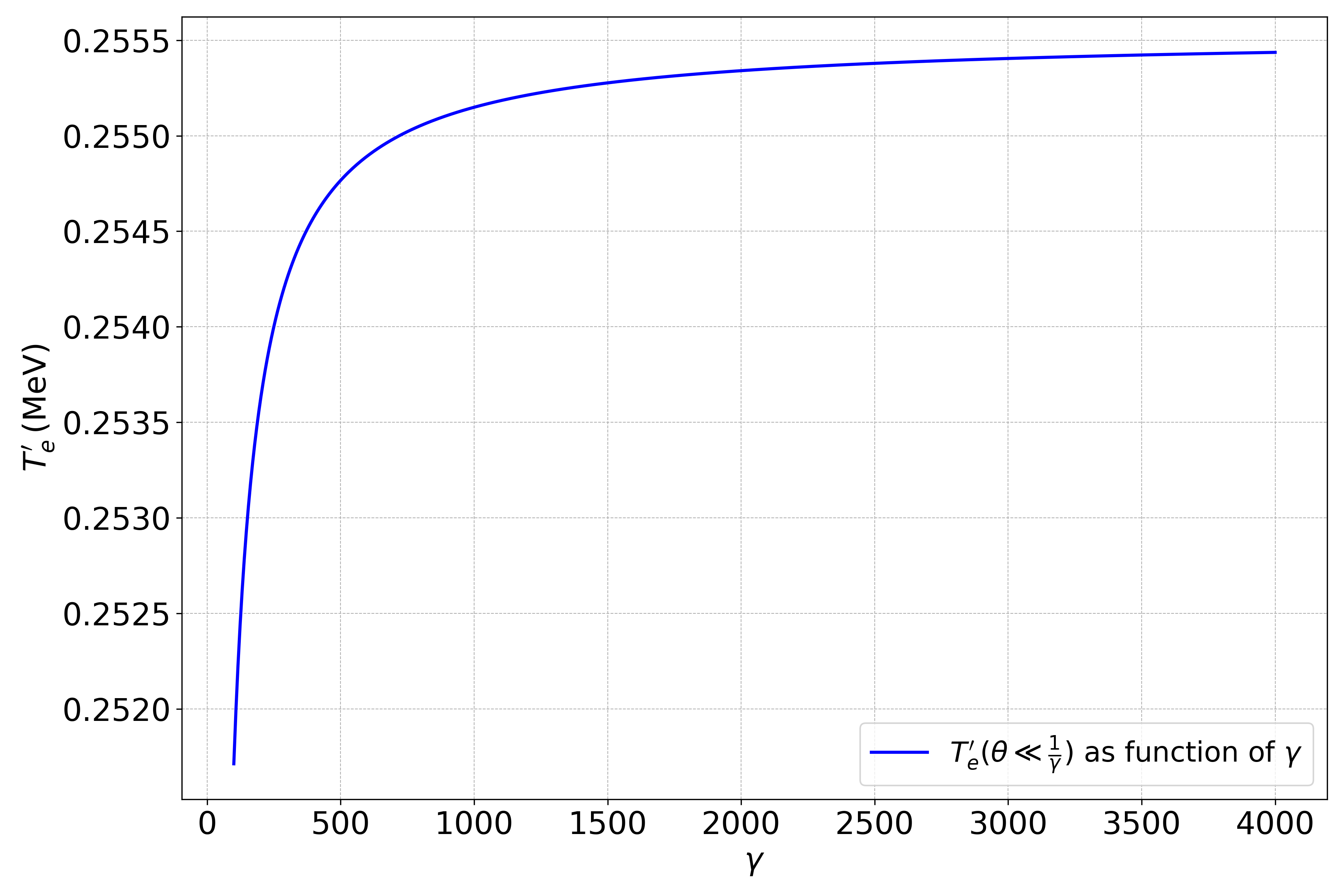}
    \caption{Kinetic energy of the scattered electron $T_{e}^{\prime}(\theta=1/\gamma)$ as function of the Lorentz factor for $50 \leq \gamma \leq 4000$.}
    \label{fig:TeprimevsgammaB}
\end{figure}

This is the counterpart of back-scattered photons within an angle $\theta = \frac{1}{\gamma}$ to have an energy very close to the maximum one at $\theta$=0. Actually, developing Eq. \ref{eq:Eph_compton} following the calculations reported in Ref. \cite{Recoil17}, we can derive the dependence of the scattered photon energy as a function of the scattering angle $\theta$ in case of relativistic electrons ($\gamma \gg 1$) and large recoil $X\gg1$ .

%INSERIRE FORMULA   E'ph = (1-(1+gam2tet2)/X)Ee
%add paragraph for this equation E'ph = Ee (1 - 1/X - gamma^2teta^2/X)

\begin{equation}
\label{eq:Eph'_X}
    E_{ph}^{\prime} = E_e \: (1 - (\frac{1+\gamma^2 \theta^2}{X}))
\end{equation}

The energy difference between photons scattered at $\theta$=0 and those scattered at $\theta = \frac{1}{\gamma}$ is minimal, of the order of $\frac{1}{X}$. This must be compared to the Thomson regime where instead photons scattered at an angle $\theta \neq 0$ have an energy given by $E_{ph}^{\prime}$= $4 \gamma^2 E_{ph}/(1+\gamma^2 \theta^2)$ . Consequently, at $\theta = \frac{1}{\gamma}$ the photon energy is half of its maximum value at $\theta$=0.

In analyzing the momentum of the electron post-collision, we completed the characterization of the scattered electrons for FICS in terms of both kinetic energy and momentum components. For simplicity, we consider the scattering to occur in the x-plane. When the photon scatters at a small angle $\theta$, its momentum in the x-direction is approximately $\theta E_e$ (or $\theta \gamma m_e c^2$), as the photon essentially absorbs the energy of the electron $E_e$.
The momentum components $p_e^{\prime}$ of the electron, in the x and z directions, are expressed as:

\begin{equation}
\label{eq:pe'x}
    p_{ex}^{\prime} = m_e c\: \gamma \theta
\end{equation}

\begin{equation}
\label{eq:pe'z}
    p_{ez}^{\prime} = m_e c \:\gamma^2 \frac{\theta^2}{2}
\end{equation}
The total momentum $p_{e}^{\prime}$ of the electron can be calculated using:
\begin{equation}
\label{eq:pe'tot}
    p_{e}^{\prime} = m_e c \:\gamma \theta \sqrt{1 + \frac{\theta^2}{4}}
\end{equation}

We can now compute the FICS fraction factor denoted $f^{FICS}=\frac{\mathscr{N}^\Psi}{\mathscr{N}}$. This factor represents the proportion of electrons that undergo FICS, specifically those that correspond to photons scattering within the angle $\theta=\frac{1}{\gamma}$.
To derive this factor, we consider the number of photons within a normalized acceptance angle $\Psi$. The detailed formulation and derivation of this factor are thoroughly presented in Ref. \cite{PhysRevAccelBeams.20.080701}, particularly in Eqs. 19 and 20. These equations provide the mathematical framework necessary to quantify the number of photons in the acceptance angle, thereby enabling the calculation of the FICS fraction factor as follow:

\begin{equation}
\label{eq:ffics}
    f^{FICS} = \frac{6.25}{4.2} \frac{(1+ \sqrt[3]{X}\:\frac{\Psi^2}{3}) \Psi^2}{(1+ (1+\frac{X}{2}) \Psi^2)(1+\Psi^2)} \cdot \frac{\sigma_T}{\sigma}
\end{equation}

where $\Psi \approx \frac{\gamma\theta}{\sqrt{1+X}}$ is the normalized acceptance angle \cite{4599118} in its generalized form accounting for recoil effects, $\sigma$ is the unpolarized Compton cross section and $\sigma_T$ is the Thomson cross section \cite{Berestetskii:1982qgu}.

Taking the fact that $\frac{\sigma_T}{\sigma}=\frac{4X}{lnX+\frac{1}{2}}$, $X\gg 1$, $\gamma \gg 1$ and $\theta \le \frac{1}{\gamma}$, Eq.\ref{eq:ffics} simplifies to the following form:

\begin{equation}
\label{eq:ffics_simp}
    f^{FICS}(\theta=\frac{1}{\gamma}) \approx \frac{4}{ln X+\frac{1}{2}}
\end{equation}

Taking into consideration that the recoil $X=\frac{4E_eE_{ph}}{(m_0c^2)^2}$ for FICS simplifies to $2\gamma$ (see Tab. 1 in Ref. \cite{serafiniFICS}). Eq. \ref{eq:ffics_simp} can be written as:

\begin{equation}
\label{eq:ffics_gamma}
    f^{FICS}(X=2\gamma) = \frac{4}{\mathrm{ln} \gamma+1.19}
\end{equation}

\begin{figure}[!ht]
    \centering
    \includegraphics[width=\columnwidth]{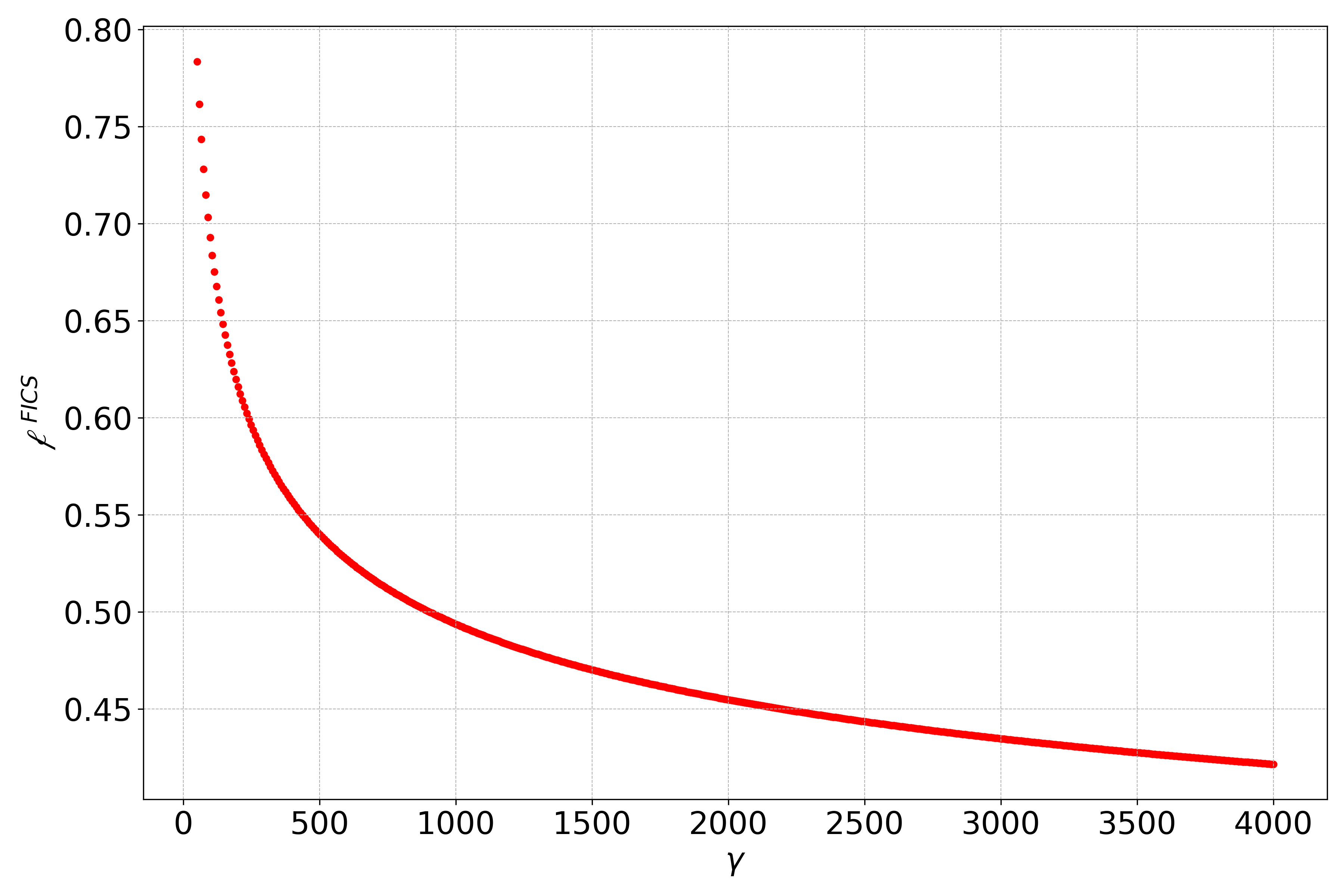}
    \caption{FICS fraction factor $f^{FICS}$ as function of the Lorentz factor $\gamma$.}
    \label{fig:factorf}
\end{figure}

Fig. \ref{fig:factorf} clearly shows that the fraction of electrons which are stopped to quasi-stationary conditions after undergoing FICS, with maximum kinetic energy of 255.5 keV and maximum relative speed of $\beta$=0.75, are a significant fraction of the total number of scattering events, going from almost 70$\%$ at lower gammas down to 40$\%$ for $\gamma$ = 4000. The logarithm behavior shown by Eq. \ref{eq:ffics_gamma} states that the fraction factor $f^{FICS}$ remains significant even for very large values of $\gamma$.

\section{Computation of Full Inverse Compton Spectra in the Linear Regime}

In this study, we employed the WHIZARD code to compute the spectra of FICS interaction between a photon and an electron in the linear regime. WHIZARD is a powerful and versatile software tool designed for simulating high-energy particle interactions \cite{Kilian2011}. It is particularly adapt at handling complex scattering processes and providing detailed spectral outputs. We analyzed the interaction of a 255.5 keV ($m_ec^2/2$) photon with a 200 MeV electron as an initial computational step. The code's precise modeling capabilities enabled us to take into account various physical parameters and interaction regimes, making it an ideal tool for this study. The results demonstrated the stopping power of 255.5 keV photon energy in halting any ultra-energetic electron in a single collision event, providing valuable insights into the dynamics of FICS and contributing to a deeper understanding of the underlying physics in high-energy particle collisions.

\begin{figure}[!ht]
    \centering
    \includegraphics[width=\columnwidth]{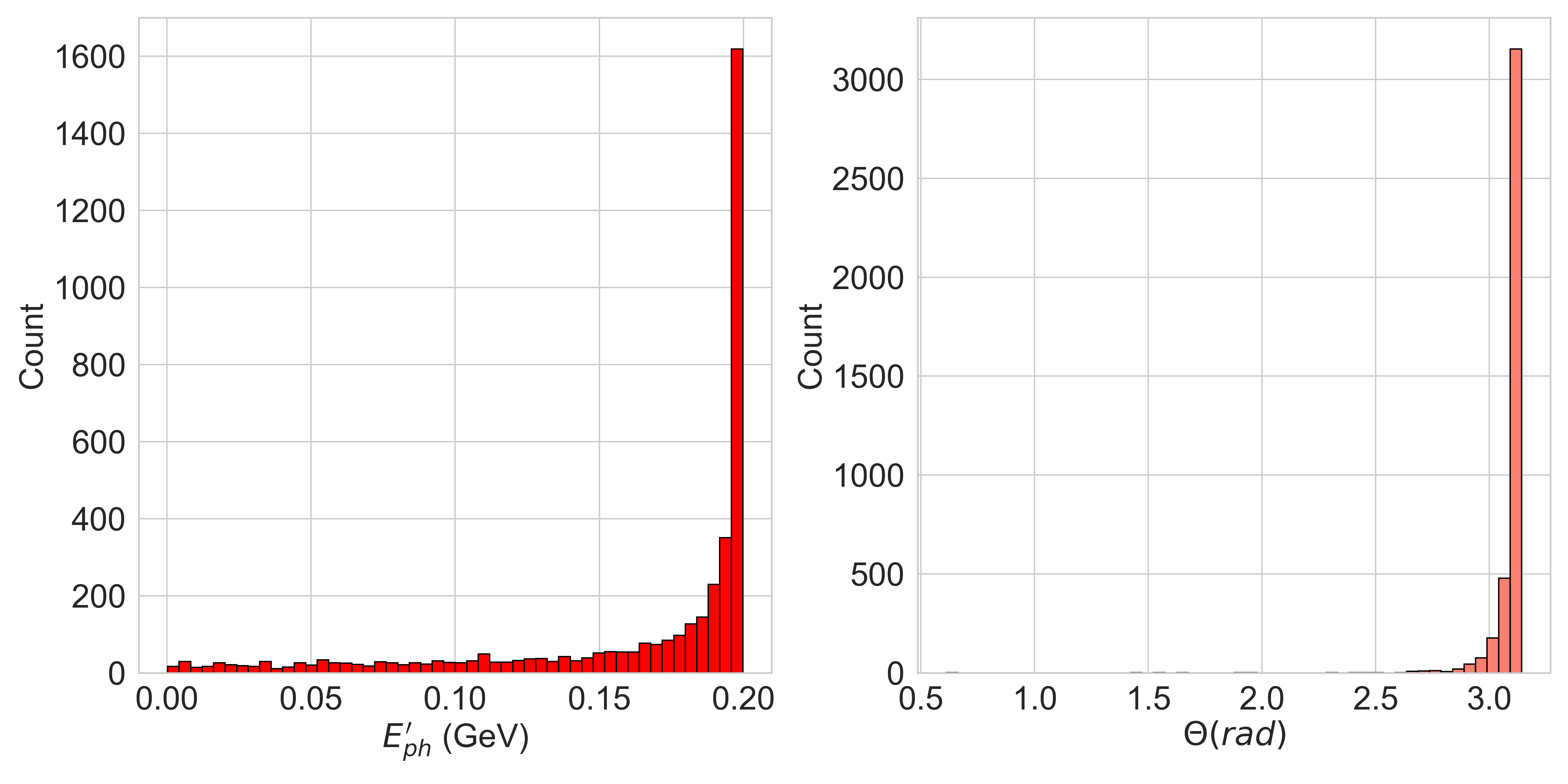}
    \caption{Energy Spectrum and Scattering Angle Distribution of Scattered Photons after FICS interaction. Left Panel: energy spectrum of scattered photons after doing FICS. The x-axis represents the energy in GeV, while the y-axis indicates the counts. Right Panel: scattering angle distribution of the scattered photons.}
    \label{fig:Eph_wizard}
\end{figure}

The spectral distribution the energy of the scattered photons after FICS is shown in the left diagram of Fig. \ref{fig:Eph_wizard}, while the angular distribution is in the right diagram. Most of the photons scatter at an energy close to 200 MeV, which can be seen as well in Fig. \ref{fig:Pz_ph} that corresponds to the incident electron energy and they are back-scattered at an angle close to zero, which is the direction of propagation of incident electrons. 

On the other hand, Fig. \ref{fig:Eph&events} clearly illustrates that the analytical formula derived in Eq. \ref{eq:Eph'_X}, which calculates the energy of scattered photons, is in perfect alignment with the numerical results obtained. The consistency between the theoretical predictions and the empirical data not only validates the formula but also reinforces the robustness of the underlying physical principles that govern photon scattering in this context of FICS.

\begin{figure}[!ht]
    \centering
    \includegraphics[width=\columnwidth]{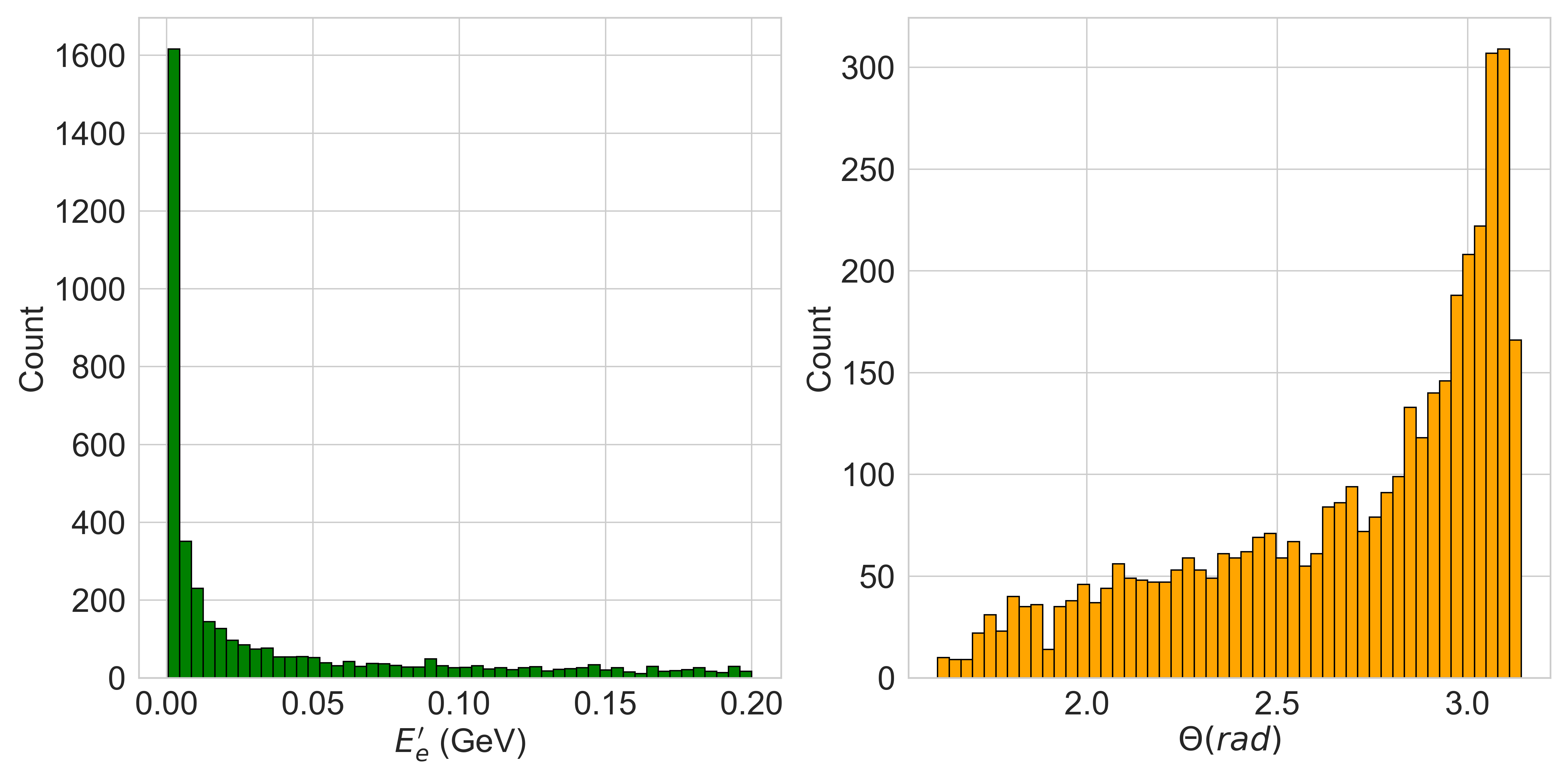}
    \caption{Simulations of the energy spectrum and scattering angle distribution of scattered electrons following FICS for $E_e=200$ MeV and $E_{ph}=255.5$ keV. Left Panel: energy spectrum of scattered electrons after FICS interaction. The x-axis represents the energy in GeV, while the y-axis indicates the count. Right Panel: scattering angle distribution of the scattered electrons.}
    \label{fig:Ee_wizard}
\end{figure}

\begin{figure}[!ht]
    \centering
    \includegraphics[width=\columnwidth]{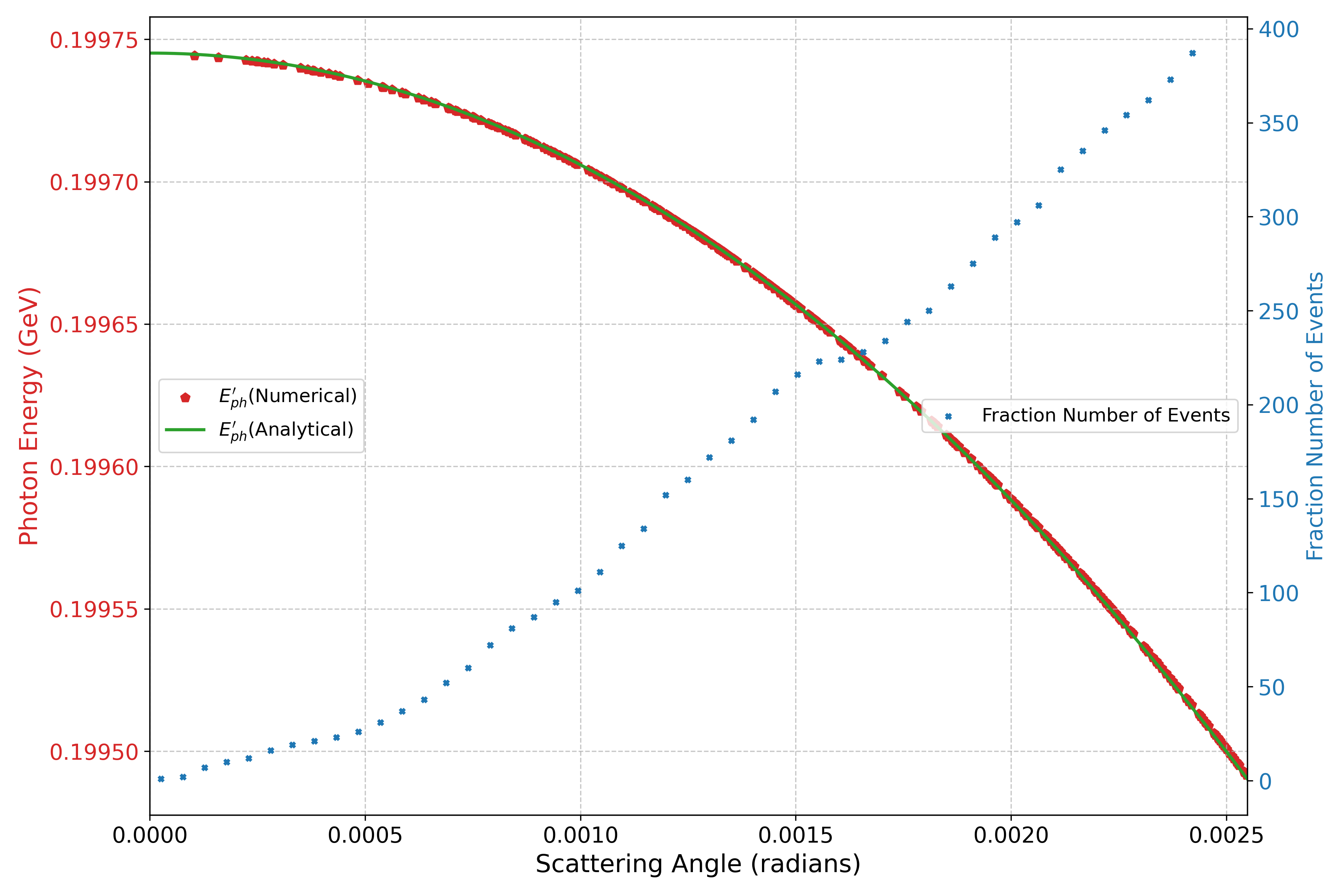}
    \caption{Comparison of Analytical (in green) and Numerical Photon Energy (in red) as functions of the scattering angle $\theta$ in the range from $\theta$=0 up to $\theta=1/\gamma$. Blue dots mark instead the number of events (right vertical axis) occuring for photons scattered within $\theta$, considering a total number of events equal to 4000. }
    \label{fig:Eph&events}
\end{figure}

\begin{figure}[!ht]
    \centering
    \includegraphics[width=\columnwidth]{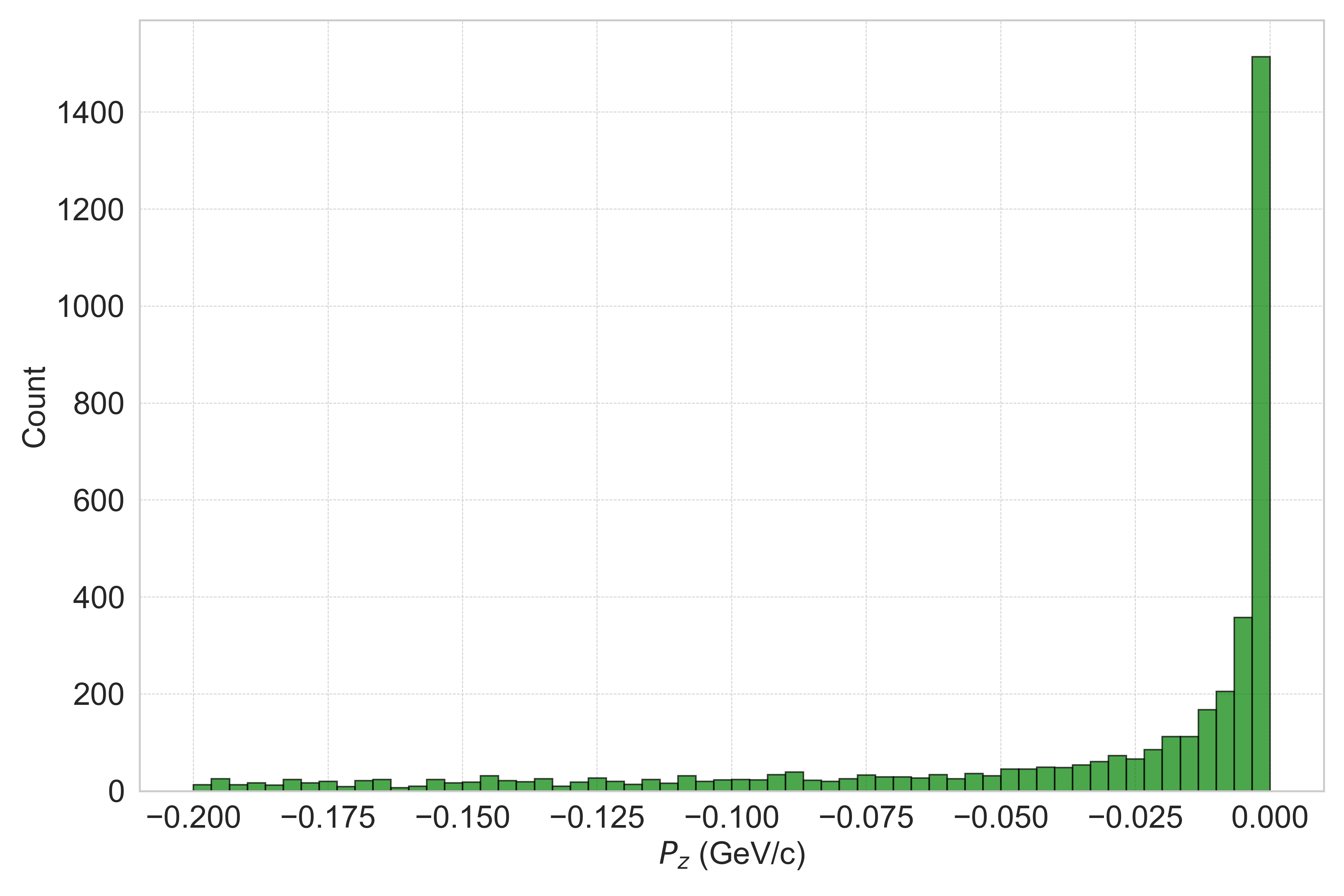}
    \caption{Distribution of the longitudinal momentum of the scattered Electrons from right to left with an initial energy of $E_e$=200 MeV.}
    \label{fig:Pz_EE}
\end{figure}

\begin{figure}[!ht]
    \centering
    \includegraphics[width=\columnwidth]{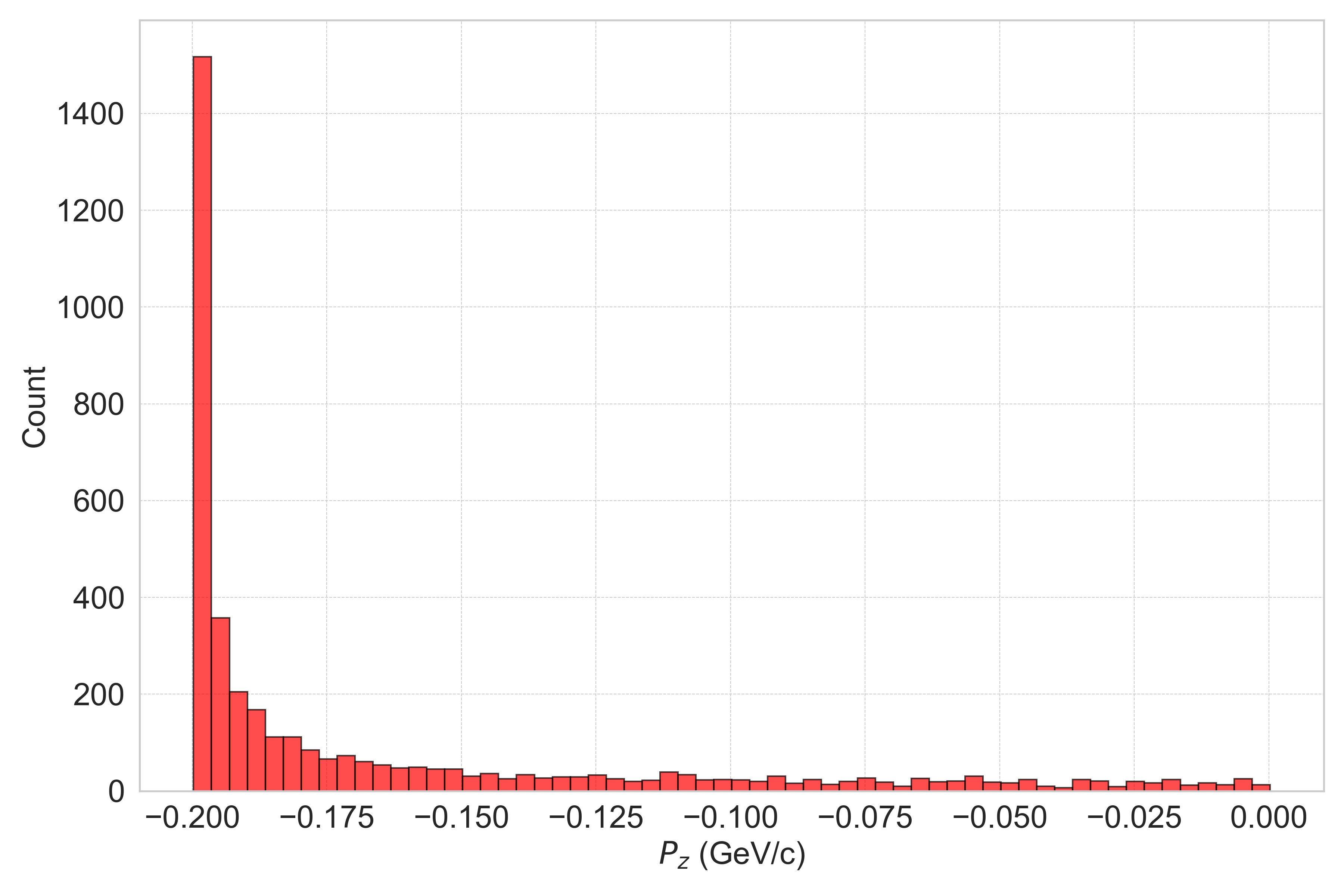}
    \caption{Distribution of the longitudinal momentum of the scattered Photons of an initial energy $E_{ph}$=255.5 keV and $E_e$= 200 MeV.}
    \label{fig:Pz_ph}
\end{figure}

Regarding the scattered electrons, their energy spectrum after fully colliding with 255.5 keV photons is illustrated in the left diagram of Fig. \ref{fig:Ee_wizard}, while the corresponding angular distribution is shown in the right diagram. The energy spectrum reveals that most of the scattered electrons are at rest, which is consistent with our theoretical results presented in previous sections. This observation aligns with the momentum distribution of the electrons depicted in Fig. \ref{fig:Pz_EE}, where a significant number of electrons have nearly zero longitudinal momentum.

\begin{figure}[!ht]
    \centering
    \includegraphics[width=\columnwidth]{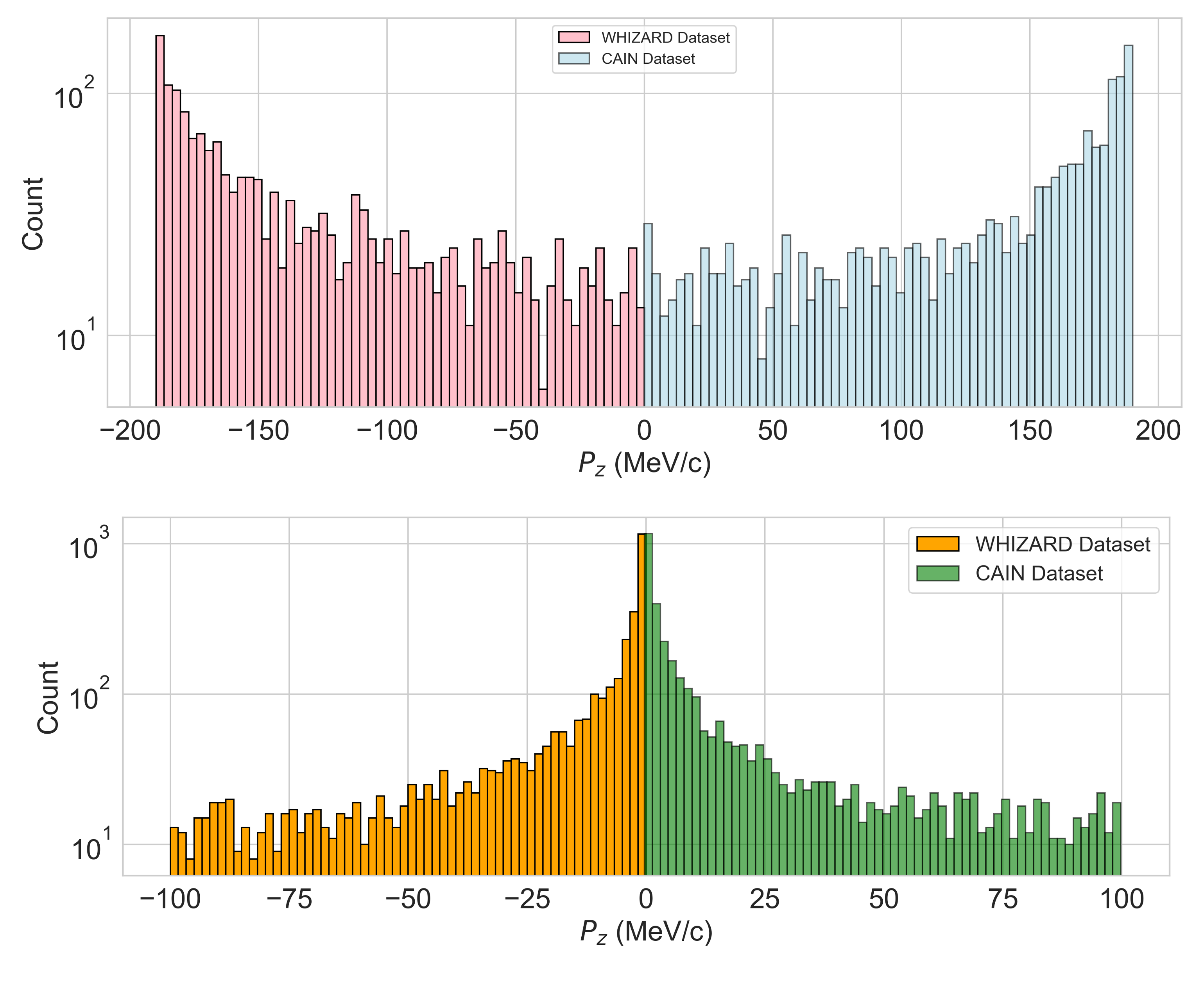}
    \caption{Momentum Spectrum ($P_z$) in MeV/c of scattered photons and electrons following FICS Interaction. Top Panel: The momentum spectra of the scattered photons of an initial energy $E_{ph}$=255.5 keV, comparing simulations from CAIN (blue) and WHIZARD (pink). The x-axis represents the longitudinal momentum ($P_z$) in MeV/c, and the y-axis represents the count on a logarithmic scale. Bottom Panel: The momentum spectra of scattered electrons on an initial energy $E_e$=200 MeV, with simulations from CAIN (green) and WHIZARD (orange).}
    \label{fig:pz_w_c}
\end{figure}

In Fig. \ref{fig:pz_w_c} we compare the longitudinal momentum $(P_z)$ distributions of the scattered photons and electrons after a full inverse scattering interaction, simulated using both WHIZARD and CAIN codes \cite{CAIN}. The top plot shows $P_z$ of scattered photons from an interaction where electrons, initially at 200 MeV, lose their kinetic energy to the photons. Both simulations display similar distributions: note that electrons in WHIZARD scatter from right to left (opposite convention than in CAIN, where they scatter from left to right). The bottom plot illustrates the momentum of scattered electrons that transferred all their kinetic energy to photons with an initial energy of 255.5 keV. The results from WHIZARD and CAIN again show a similar pattern, reinforcing the consistency between the two codes. The logarithmic scale on the y-axis highlights this consistency across different momentum magnitudes. This agreement validates the reliability of both codes in simulating FICS interactions, demonstrating their effectiveness in modeling the physics of these interactions and their suitability for accelerator physics studies.

\begin{figure}[!ht]
    \centering
    \includegraphics[width=\columnwidth]{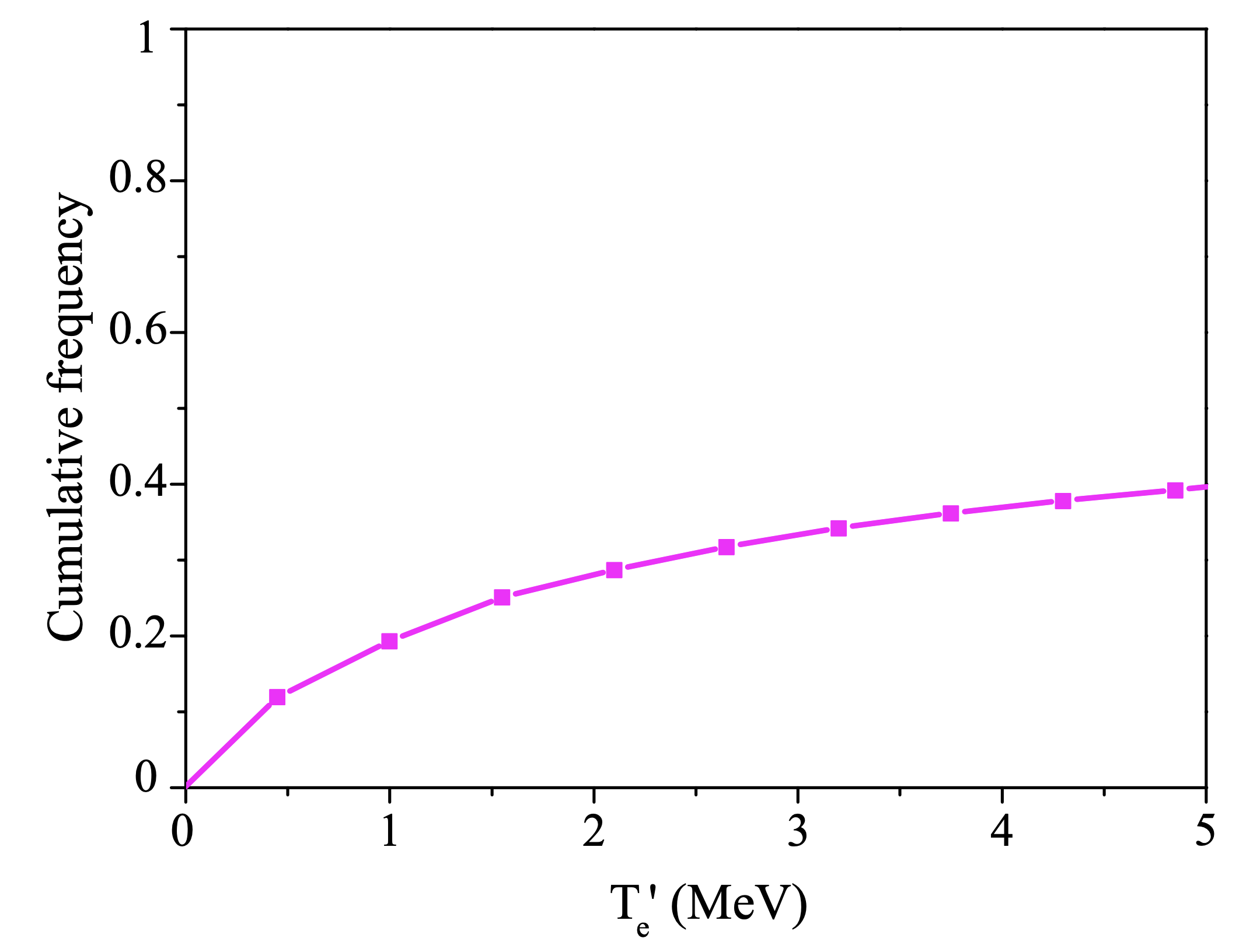}
    \caption{Relative fraction of events as function of the kinetic energy of the scattered electrons $T_e^{\prime}$ as specified in x-axis.}
    \label{fig:Ee&events}
\end{figure}

In Fig.14 we show an important result obtained analyzing output data from WHIZARD that is complementary to those shown in Fig.10: the relative fraction of electrons undergoing scattering and taken down to small low-relativistic velocities by the FICS interaction. The relative fraction of scattered electrons over the total amount of events is shown on the vertical axis as function of the maximum kinetic energy of the scattered electrons (in MeV). Here the first data point shows that almost 12$\%$ of the scattered electrons have a kinetic energy smaller than 490 keV, which corresponds to a velocity equal to 0.86$\times$c. This is of relevance for the discussion reported in the next section concerning the capability of FICS to stop electrons strongly reducing their initial velocity (that is almost equal to c for the electrons energy considered here, in excess of tens of MeV) down to a fraction of the speed of light, therefore generating very large negative accelerations.

\section{Unruh radiation}

The electrons performing Compton scattering in the FICS condition undergo a state transition with a longitudinal momentum change by a quantity almost equal to its initial value. Their velocity, in turn, changes by a factor $\Delta v \simeq v_0$, $v_0\simeq c$ being the electron velocity before the scattering. The typical time duration of this transition can be evaluated by applying the Tamm-Mandelstam criterion \cite{Tamm}:

\begin{equation}
    \Delta E \Delta t \simeq h
\end{equation}
In the cases presented previously,
\begin{equation}
    \Delta E = E_e-E'_e \simeq E_e-m_ec^2
\end{equation}
and the deceleration time turns out to be:
\begin{equation}
    \Delta t \simeq \frac{h}{E_e-m_ec^2} 
\end{equation}

leading to:

\begin{equation}
    a \simeq \frac{c\: (E_e-m_ec^2)} {h}
\end{equation}

\begin{figure}[ht!]
    \centering
    \includegraphics[width=\columnwidth]{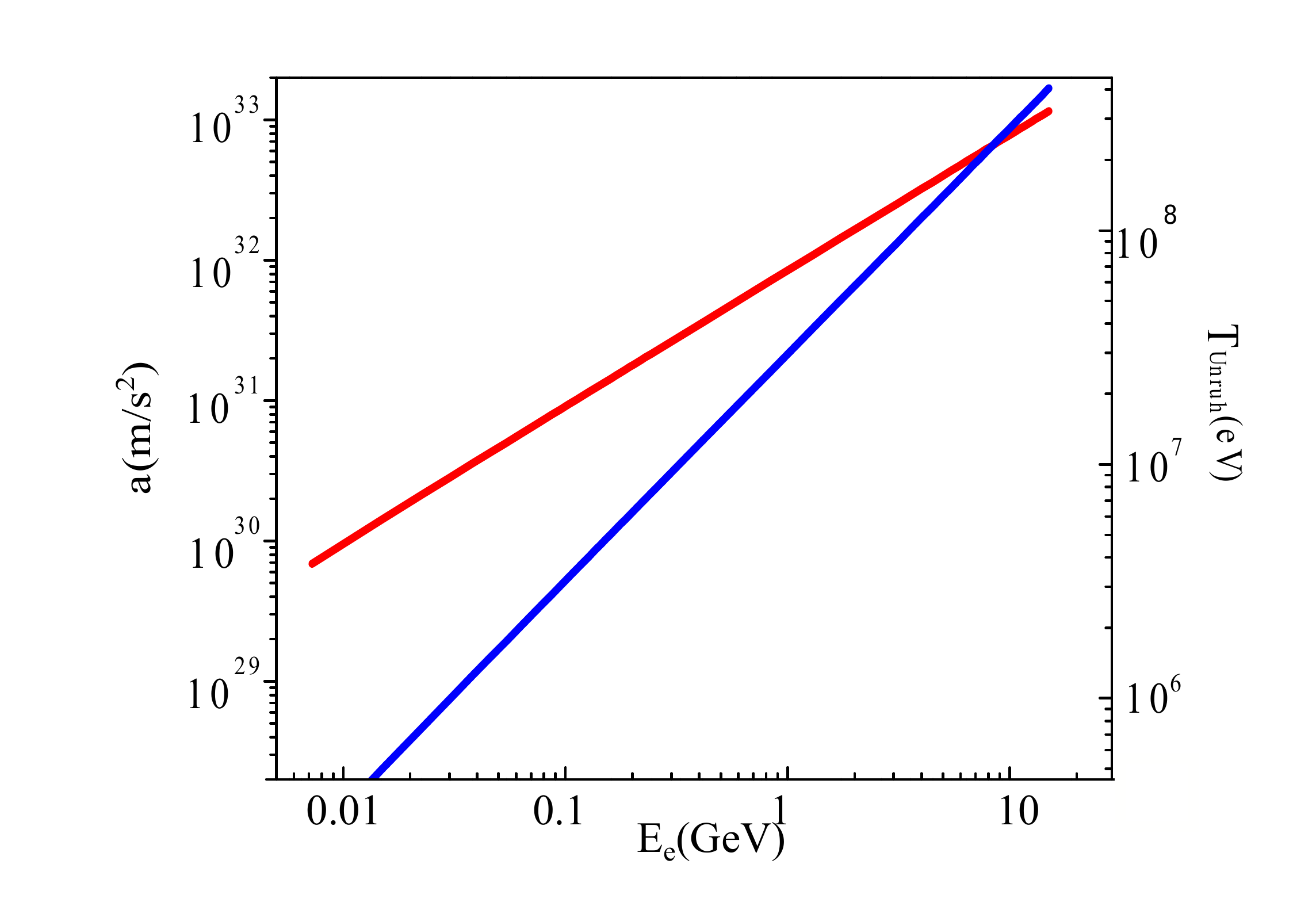}
    \caption{Acceleration in $m/s^2$ of the electrons (red) and temperature in eV (blue) of the Unruh thermal bath as function of the initial electron energy (GeV)}
    \label{fig:Unruh}
\end{figure}

Fig. \ref{fig:Unruh} shows the considerable values that the acceleration could attain as a function of the electron energy.
Undergoing an acceleration can be assimilated to experiencing the Minkowski vacuum as a thermal bath with the Unruh temperature \cite{Unruh2}. 
$T_{Unruh} =\frac{\hbar a}{2\pi k_B c}$, which has been reported in Fig. \ref{fig:Unruh} on the right axis. 
The accelerated electron behaves as if it was immersed in such a thermal bath and there is therefore a non-zero probability that
it absorbs a virtual particle from this bath and passes to an excited state, performing Compton Scattering. Seen in the laboratory,
this process corresponds to the emission of a real photon [5]. Conversely, also the process of re-emission by the electron of
the virtual particle into the bath in the accelerated frame, going back to its ground state, corresponds to the emission of a real photon in the inertial frame.
When the time between absorption and re-emission is small, the electron scatters the virtual photons of the thermal bath.
This process, corresponds to the emission of two real particles in the laboratory by the accelerated scatterer and could be considered as a signature of the Unruh effect. Fig. \ref{fig:Unruh} clearly shows that acceleration levels attainable with FICS are several orders of magnitude larger than those discussed in Refs. \cite{unruh17} and \cite{RevModPhys.77.1131} using interactions of ultra-relativistic electrons with either high intensity lasers or channeling effects in crystals.

\section{Conclusion}

We showed characteristics of a specific Compton scattering modality that assures the complete transfer of energy and momentum between the incident electron and the colliding photon. This modality named FICS.
Full Inverse Compton Scattering, is the only way to stop an electron propagating in vacuum: surprisingly enough the incident photon must have an energy of 255.5 keV, corresponding to one half of the electron
rest mass energy. Photons of such an energy can stop any relativistic electron of any energy, from MeV’s to TeV’s and beyond, taking them down to rest after collision, and acquiring all their kinetic energy. FICS is the time reversal of Compton scattering on atomic electrons of very high energetic photons, that are scattered back with an energy equal to 255.5 keV, after transferring their energy to the scattered electrons. In a collision of two primary beams of 255.5 keV photons and relativistic electrons of any energy, a significant
fraction of the total scattering events is confined to photons back-scattered within an angle $\theta=1/\gamma$, and their corresponding scattered electrons after collisions are taken down to kinetic energies smaller than $0.5 \times mc^2$. This implies that most of the scattering events leave electrons after collision down to almost at rest in the lab frame: the negative acceleration that electrons experience is ultra-high, being subject to stop from the speed of light to almost 0 in a time as fast as the emitted photon inverse frequency. These accelerations should make the stopping electron sense the Unruh radiation bath/temperature, opening a way to investigate this quantum gravity phenomenon using moderate and compact accelerators capable to arrange collisions between GeV-class electron beams and 255.5 keV photon beams.

\bibliographystyle{elsarticle-num-names} 
\bibliography{references.bib}


%\begin{thebibliography}{00}



%\end{thebibliography}
\end{document}